\documentclass{article}       
\usepackage{e}                
\usepackage{latexsym}
\usepackage{hyperref}
\input boxedeps
\SetepsfEPSFSpecial 
\HideDisplacementBoxes

\newcommand{\etal}{{\em et al.}}
\newcommand{\ie}{{\em i.e.}}

\newcommand{\gevcc}{\hbox{ GeV}\!/\!c^2}
\newcommand{\gev}{\hbox{ GeV}}

\newcommand{\tevcc}{\hbox{ TeV}\!/\!c^2}

\newcommand{\mm}{\hbox{ mm}}

\newcommand{\m}{\hbox{ m}}
\newcommand{\lum}{\hbox{ cm}^{-2}\hbox{ s}^{-1}}
\newcommand{\eqn}[1]{(\ref{#1})}
\def\url#1{\mbox{\href{#1}{\sf #1}}}

\def\ltap{\mathop{\raisebox{-.4ex}{\rlap{$\sim$}} 
\raisebox{.4ex}{$<$}}}
\def\gtap{\mathop{\raisebox{-.4ex}{\rlap{$\sim$}} 
\raisebox{.4ex}{$>$}}}

\def\vev#1{\langle #1\rangle_0}
\def\abs#1{\left| #1\right|}

\newcommand{\hepex}[1]{hep-ex/#1}
\newcommand{\hepph}[1]{hep-ph/#1}

\newcommand{\astro}[1]{astro-ph/#1}
\def\prl#1#2#3{\frenchspacing{\it Phys. Rev. Lett. }{\bf #1} (19#3) #2}
\def\prll#1#2#3{\frenchspacing{\it Phys. Rev. Lett. }{\bf #1} (#3) #2}
\def\pr#1#2#3{\frenchspacing{\it Phys. Rev. D}{\bf #1} (19#3) #2}
\def\prM#1#2#3{\frenchspacing{\it Phys. Rev. D}{\bf #1} (#3) #2}

\def\pl#1#2#3{\frenchspacing{\it Phys. Lett. }{\bf #1} (19#3) #2}
\def\np#1#2#3{\frenchspacing{\it Nucl. Phys. }{\bf #1} (19#3) #2}
\def\npM#1#2#3{\frenchspacing{\it Nucl. Phys. }{\bf #1} (#3) #2}
\def\rmp#1#2#3{\frenchspacing{\it Rev. Mod. Phys. }{\bf #1} (19#3) #2}

\def\app#1#2#3{\frenchspacing{\it Acta Phys. Polon. B}{\bf #1} (19#3) #2}

\def\phystoday#1#2#3#4{\frenchspacing{\it Phys. Today\/ }{\bf #1}
(\ifcase#3\or January\or 
         February\or March\or April\or May\or June\or July\or August\or 
         September\or October\or November\or December\fi, 19#4) #2}

\begin{document}
%
%
\thispagestyle{empty}
\branch{C}   
%
\title{Visions:  The Coming Revolutions  \\  in Particle Physics}
\author{Chris Quigg}
\institute{Fermi National Accelerator Laboratory,\thanks{Fermilab is
operated by Universities Research Association Inc. under Contract No. 
DE-AC02-76CH03000 with the United States Department of Energy.} P.O.
Box 500, Batavia, Illinois 60510 USA; e-mail: quigg@fnal.gov}
\PACS{12.15.-y 13.85.-t 12.60.-i 14.80.Bn \hfill \textsf{FERMILAB--CONF-02/058--T}}
\maketitle
\begin{abstract}
Wonderful opportunities await particle physics over the next decade,
with the coming of the Large Hadron Collider to explore the 1-TeV scale
(extending efforts at LEP and the Tevatron to unravel the nature of
electroweak symmetry breaking) and many initiatives to develop our
understanding of the problem of identity and the dimensionality of
spacetime.  
\end{abstract}
\section{Galileo's Three Revolutions}
In this talk, I want to evoke some of the revolutionary developments I
believe will come over the next two decades in particle physics. To set
the context, and because we find ourselves in Italy, I'd like to begin
by recalling three revolutions we identify with Galileo and his time.

\noindent\textit{Eppur si muove.} In the public mind and in popular
literature, Galileo is remembered chiefly for his part in completing
the Copernican revolution, establishing that humans do not occupy a
privileged location in the universe. It's a great achievement and a
good story, given texture by Galileo's complex persona and by the
richness of his relationship with the inquistorial Church~\cite{dava}.

\noindent\textit{Cimenti.} The Copernican revolution is a scientific
movement accomplished. We scientists revere Galileo no less for his
contribution to the scientific method. For it was during Galileo's time
that humans found the courage to reject Authority. They learned instead
to read nature by doing experiments, subjecting their hypotheses to
unremitting trials by ordeal that Galileo called \textit{cimenti.} 
The notion that experiment, not eloquence, is the arbiter of what is 
true revolutionized mankind's relationship with nature.

\noindent\textit{The minute particular.} An essential element of
civilization is human curiosity about the world and a
thirst to comprehend nature. Until five centuries ago, the questions
our ancestors wondered about were broad and the explanations they
advanced were sweeping but vague. Asking great questions---seeking to
explain everything about the world all at once---led to extremely
limited answers. Science as we know it took shape in Galileo's time
when humans learned that asking limited questions could lead them to
universal insights. In contemporary American discourse at least, the
received wisdom holds that all the great questions have been answered
and that today's scientists---gifted though they might be---are dealing
with ever narrower research topics. This canard betrays an ignorance of
how science has been done ever since it became worthy of the name.
You can help explain to the world what science is. To be most
effective, you must make the connection between your own minute particular
(the search for the Higgs boson, the width of the $W$, the mass of the
top, or whatever) and the universal understanding we are trying to
build.\footnote{A recent article in \textit{FermiNews}~\cite{fnews}
makes the challenge uncomfortably plain.}

Are all the great scientific revolutions in the distant past? Not at
all: We are here to discuss the revolution-in-progress that we expect
experiments with the Large Hadron Collider to complete. This is the
radically new and simple conception of matter brought about
by the development of gauge theories and the recognition that quarks and
leptons are the basic constituents of matter---at the current limits of
our resolution.  The gauge-theory synthesis is part of a larger change that
we are living through---no, making---in the way humans think about their
world. The recognition that the human scale is not privileged, that we need
to leave our familiar surroundings the better to understand them, has been
building since the birth of quantum mechanics. As it emerges whole, fully
formed, in our unified theories and renormalization group equations, the
notion seems to me both profound and irresistible. I find it fully appropriate
to compare this change in perception with the shifts in viewpoint we owe to
Copernicus and Einstein.

\section{Our Picture of Matter}
We base our understanding of physical phenomena on the identification
of a few constituents that seem elementary at the current limits of
resolution of about $10^{-18}\m$, and a few fundamental forces.  The
constituents are the pointlike quarks \{$(u, d)_{L}$, $(c, s)_{L}$, $(t,
b)_{L}$\} and leptons \{$(\nu_{e}, e)_{L}$, $(\nu_{\mu}, \mu)_{L}$,
$(\nu_{\tau}, \tau)_{L}$\}, with strong, weak, and electromagnetic
interactions specified by $SU(3)_{c}\otimes SU(2)_{L}\otimes U(1)_{Y}$
gauge symmetries.

The electroweak theory is founded on the weak-isospin symmetry 
embodied in the quark and lepton doublets
and weak-hypercharge phase symmetry, plus the idealization that 
neutrinos are massless.\footnote{For surveys of the electroweak
theory, with references, see Ref.\
\cite{Quigg:1999xg,Quigg:2001td}.} In its simplest form, with the
electroweak gauge symmetry broken by the Higgs mechanism, the
$SU(2)_{L}\otimes U(1)_{Y}$ theory has scored many qualitative
successes: the prediction of neutral-current interactions, the
necessity of charm, the prediction of the existence and properties of
the weak bosons $W^{\pm}$ and $Z^{0}$.  Over the past ten years, in
great measure due to the beautiful experiments carried out at the $Z$
factories at CERN and SLAC, precision measurements have tested the
electroweak theory as a quantum field theory, at the one-per-mille
level~\cite{Sirlin:1999zc,Swartz:1999xv,Charlton:2001am}.
Last year, our colleagues working at LEP made a heroic push to discover 
the Higgs boson~\cite{unknown:2001xw}.  The search will intensify 
again in a few years at the Tevatron and the Large Hadron Collider.

The quark model of hadron structure and the parton model of 
hard-scattering processes have such pervasive influence on the way we 
conceptualize particle physics that quantum chromodynamics, the theory 
of strong interactions that underlies both, sometimes is taken for granted.  
QCD is a remarkably simple, successful~\cite{Quigg:1999di}, and rich 
theory of the strong interactions~\cite{Wilczek:1999id}.  The 
perturbative regime of QCD exists, thanks to the crucial property of 
asymptotic freedom, and describes many phenomena in quantitative 
detail.  The strong-coupling regime controls hadron structure and 
gives us our best information about quark masses.  Unfamiliar
r\'{e}gimes of high density and high temperature contain riches we
have only begun to explore.

The physics curriculum in the 1898--99 University of 
Chicago catalogue begins with a very triumphalist Victorian preface~\cite{sbt}:
\begin{quote}
    ``While it is never safe to affirm that the future of the Physical
    Sciences has no marvels in store even more astonishing than those of
    the past, it seems probable that most of the grand underlying
    principles have been firmly established and that further advances are
    to be sought chiefly in the rigorous application of these principles
    to all the phenomena which come under our notice \ldots .  An eminent
    physicist has remarked that the future truths of Physical Science are
    to be looked for in the sixth place of decimals.''
\end{quote}
As the ink was drying on these earnest words, R\"{o}ntgen discovered x
rays and published the epoch-making radiograph of his wife's hand,
Becquerel and the Curies explored radioactivity, Thomson discovered
the electron and showed that the ``uncuttable'' atom had parts, and
Planck noted that anomalies in the \textit{first} place of the
decimals required a wholesale revision of the physicist's conception of
the laws of Nature.

We have the benefit of a century of additional experience and insight,
but we are not nearly so confident as our illustrious Victorian ancestors were
 that we have uncovered ``most of
the grand underlying principles.''  Indeed, while we celebrate the
insights codified in the \textit{standard model of particle physics}
and look forward to resolving its puzzles, we are increasingly
conscious of how little of the physical universe we have experienced and explored. 
Future truths are still to be found in precision measurements, but the
century we are leaving has repeatedly shown that Nature's marvels are
not limited by our imagination.  Exploration can yield surprises that
completely change what we think about---and how we think.

\subsection{Precision Measurements of Electroweak Observables}
The beautiful agreement between the electroweak theory and a vast 
array of data from neutrino interactions, hadron collisions, and 
electron-positron annihilations at the $Z^{0}$ pole and beyond means 
that electroweak studies have become a modern arena in which we can 
look for new physics ``in the sixth place of 
decimals.'' Classic achievements include  determining of the
number of light neutrino species and inferring the mass of the 
top quark.\footnote{These are reviewed in \S4.1 of my \textit{TASI~2000}
Lectures, Ref.~\cite{Quigg:2001td}.}

\begin{figure}[tb] 
\centerline{\BoxedEPSF{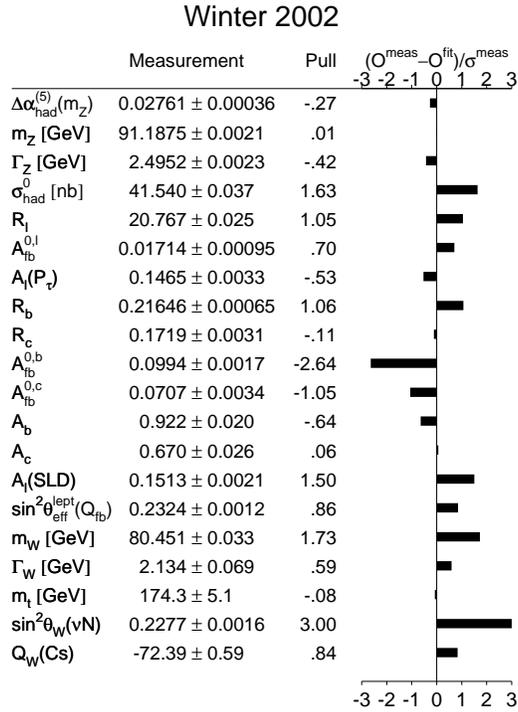 scaled 350}}
\vspace{-4pt}
\caption{Precision electroweak measurements and the pulls they exert 
on a global fit to the standard model, from Ref.\ 
{\protect\cite{ewwg}}.}
\label{fig:pulls}
\end{figure}

The comparison between the electroweak theory and a considerable 
universe of data is shown in Figure \ref{fig:pulls}~\cite{ewwg}, 
where the pull, or difference between the global fit and measured 
value in units of standard deviations, is shown for some twenty 
observables.
The distribution of pulls for this fit, due to the LEP Electroweak 
Working Group, is not noticeably different from a normal 
distribution, and only a couple of observables differ from the fit by 
as much as two standard deviations.  Whether the outliers ($A_{\mathrm{fb}}^{0,b}$ and
the NuTeV determination~\cite{Zeller:2001hh} of $\sin^2\theta_{\mathrm{W}}(\nu N)$)
contribute much-needed $\chi^2$ or point to new physics remains an open 
question~\cite{Chanowitz:2001bv}.

\subsection{Electroweak Symmetry Breaking and Masses}
Elucidating the mechanism of electroweak symmetry breaking is the most
urgent problem in particle physics, and the centerpiece of the LHC
science program. The exploration of the 1-TeV scale that will be needed
to get to the bottom of electroweak symmetry breaking is frequently
abbreviated as the search for the Higgs boson. That's a convenient
sound bite, but we mustn't let the shorthand narrow our view of the
task---and opportunity---we face. ``The search for the Higgs boson''
also fails to convey the full majesty of what we hope we will learn:
why this world of stable structures---of diversity and change---is as we
observe it to be, and not a formless fog of interchangeable parts
flying about at the speed of light.

Another sound bite is even more pernicious because it is false: that is the 
bald assertion that the Higgs boson is the source of all mass. It is not. 
There are, in fact, several kinds of mass---so several sources of mass---and we have
many interesting stories to tell.  

The masses of the hadrons are (in principle, and with increasing
precision in practice) understood from QCD in terms of the energy
stored to confine a color-singlet configuration of quarks in a small
volume~\cite{Wilczek:be,Aoki:1999yr}.  To say it another way: \textit{quantum
chromodynamics explains nearly all of the visible mass in the universe.}
This is a remarkable achievement, and we should not be hiding it.  

The Higgs boson \textit{is} an essential part of the analogy to the
Meissner effect in superconductivity that leads us to an excellent
understanding of the masses of the electroweak gauge bosons $W^{\pm}$
and $Z^{0}$ as consequences of electroweak symmetry breaking. At tree
level in the electroweak theory, we have\footnote{Although for the
moment we take the weak mixing parameter $\sin^{2}\theta_{W}$ from
experiment, we understand how it arises in a unified theory.  Moreover,
in a unified theory we can hope to understand the parameter
$\Lambda_{\mathrm{QCD}}$ that sets the scale of the hadron masses.}
\begin{eqnarray*}
    M_{W}^{2} & = & g^{2}v^{2}/2 = 
    \pi\alpha/G_{F}\sqrt{2}\sin^{2}\theta_{W} ,
    \label{eq:Wmass}  \\
    M_{Z}^{2} & = & M_{W}^{2}/\cos^{2}\theta_{W} ,
    \label{eq:Zmass}  
\end{eqnarray*}
where the electroweak scale $v = (G_{F}\sqrt{2})^{-\frac{1}{2}}
\approx 246\gev$ is set by the vacuum expectation value of the Higgs field.

Interactions with the Higgs field may also generate quark and (charged)
lepton masses; however, our understanding of these fermion masses is
considerably more primitive.  For each fermion mass, we require not
just the scale of electroweak symmetry breaking, but a distinct and
apparently arbitrary Higgs-fermion-antifermion Yukawa coupling to
reproduce the observed mass.  In the electroweak theory, the value of
each quark or charged-lepton mass is set by a new, unknown, Yukawa
coupling.  Taking the electron as a prototype, we define the
left-handed doublet and right-handed singlet
\begin{displaymath}
    \mathsf{L} = \left( 
    \begin{array}{c}
        \nu_{e}  \\
        e
    \end{array}
    \right)_{L} \; , \qquad \mathsf{R} \equiv e_{R}.
    \label{eq:elec}
\end{displaymath}
Then the Yukawa term in the electroweak Lagrangian is
\begin{displaymath}
    \mathcal{L}_{\mathrm{Yukawa}}^{(e)} = - 
    \zeta_{e}[\bar{\mathsf{R}}(\varphi^{\dagger}\mathsf{L}) + 
    (\bar{\mathsf{L}}\varphi)\mathsf{R}] \; ,
    \label{eq:eYuk}
\end{displaymath}
where $\varphi$ is the Higgs field, so that the electron mass is
$m_{e} = \zeta_{e}v/\sqrt{2}$.\footnote{Here I have taken the Yukawa coupling
$\zeta_{e}$ to be real.}
  For neutrinos, which may be their own
antiparticles, there are still more possibilities for new physics to
enter.  Inasmuch as we do not know how to calculate the fermion Yukawa
couplings $\zeta_{f}$, I believe that \textit{we should consider the
sources of all fermion masses as physics beyond the standard model.}
It is even conceivable that fermion masses arise from some entirely
distinct mechanism.

The values of the Yukawa couplings are vastly different for
different fermions: for the top quark, $\zeta_{t} \approx 1$, for the
electron $\zeta_{e} \approx 3\times 10^{-6}$, and if the neutrinos
have Dirac masses, presumably $\zeta_{\nu} \approx
10^{-10}$.\footnote{I am quoting the values of the Yukawa couplings at
a low scale typical of the masses themselves.} What accounts for the
range and values of the Yukawa couplings?  Our best hope until now has
been the suggestion from unified theories that the pattern of fermion
masses simplifies on high scales.  The classic intriguing prediction
of the $SU(5)$ unified theory involves the masses of the $b$ quark and
the $\tau$ lepton, which are degenerate at the unification point for a
simple pattern of spontaneous symmetry breaking.  The different
running of the quark and lepton masses to low scales then leads to the
prediction $m_{b} \approx 3 m_{\tau}$, in suggestive agreement with
what we know from experiment~\cite{Buras:1977yy}.

Complex Yukawa couplings are a source of \textsf{CP} violation. Consider the piece of
the Yukawa Lagrangian that gives masses to down-type quarks,
\begin{displaymath}
    - \mathcal{L}_{\mathrm{Yukawa}}^{(d)} = 
     \zeta_{ij}^{d}(\bar{\mathsf{L}}_{q_i}\varphi)\mathsf{R}_{d_j}+ 
     \zeta_{ij}^{d\; *}\bar{\mathsf{R}}_{d_j}(\varphi^{\dagger}\mathsf{L}_{d_i})\; ,
    \label{eq:dYuk}
\end{displaymath}
where the two terms are related by \textsf{CP} interchange. If the
Yukawa coupling is not a real number, \ie, if $\zeta_{ij}^{d} \neq
\zeta_{ij}^{d\; *}$, then \textsf{CP} is not a symmetry of
$\mathcal{L}_{\mathrm{Yukawa}}^{(d)}$. According to my convention that the
sources of all fermion masses relate to physics beyond the standard model, the
\textsf{CP} violation that we catalogue ``within the standard model,'' which is 
to say arising from the quark mixing matrix (hence, the Yukawa couplings), is
itself a window on physics beyond the standard model. Can we learn to read it
for clues?

\subsection{$\nu$ Oscillation News}
The science that grew into particle physics began with found
beams---the emanations from naturally occurring radioactive substance
and the cosmic rays---and found beams still provide us with important
windows on the universe.  One of the great scientific detective
stories of the recent past is the developing case for neutrino
oscillations: the evidence that neutrinos produced as one flavor
($\nu_{e}$, $\nu_{\mu}$, or $\nu_{\tau}$) actually morph into other
flavors.  Long known as a theoretical possibility, \textit{neutrino
oscillation} is now all but established by the Super-Kamiokande 
experiment's observation of an up-down asymmetry in the flux of muon 
neutrinos produced by the interaction of cosmic rays in the 
atmosphere~\cite{Toshito:2001dk}.  By far the most graceful interpretation 
is that muon neutrinos produced on the far side of the Earth oscillate 
during flight in significant numbers into tau neutrinos.

The Sudbury Neutrino Observatory has added an 
important new element to our understanding of the longstanding puzzle 
of the solar neutrino deficit~\cite{Ahmad:2001an}.  SNO reports an 
impressively precise measurement of the solar neutrino charged-current 
cross section on the heavy-water that serves as their 
target-Cherenkov detector.  The measured rate implies a $\nu_{e}$ flux 
\begin{displaymath}
	{\phi^{\mathrm{CC}}_{\mathrm{SNO}}(\nu_{e}) = 1.75 \pm 0.07 ^{+0.12}_{-0.11} 
 \pm 0.05 \times 10^{6}\lum}\; ,
\end{displaymath}
where the uncertainties are statistical, systematic, and theoretical.  
They have also measured the solar neutrino elastic ($\nu_{x} e$) cross 
section with limited precision, and extracted from it the flux of 
solar neutrinos of all active flavors,
\begin{displaymath}
	\phi^{\mathrm{ES}}_{\mathrm{SNO}}(\nu_{x}) = 2.39 \pm 0.34 ^{+0.16}_{-0.14} 
 \times 10^{6}\lum \; .
\end{displaymath}
The SNO experimenters are in the right place at the right time, because the 
Super-K experiment has already given a very precise measurement of 
the solar neutrino flux from elastic ($\nu_{x} e$) scattering~\cite{Fukuda:2001nj},
\begin{displaymath}
	{\phi^{\mathrm{ES}}_{\mathrm{SK}}(\nu_{x}) = 
	2.32 \pm 0.03 ^{+0.08}_{-0.07} 
 \times 10^{6}\lum}\; .
\end{displaymath}
The difference between the flux of active neutrinos and the flux of 
electron neutrinos,
\begin{displaymath}
	{\phi^{\mathrm{ES}}_{\mathrm{SK}}(\nu_{x}) -
	\phi^{\mathrm{CC}}_{\mathrm{SNO}}(\nu_{e}) = 0.57 \pm 0.17\times
	10^{6}\lum}\; ,
\end{displaymath}
demonstrates at $3.3\sigma$ that active neutrinos other than $\nu_{e}$, namely
$\nu_{\mu}$ and $\nu_{\tau}$, arrive at Earth.  Since the nuclear 
processes that power the Sun yield only $\nu_{e}$, this new result 
rules in favor of neutrino oscillations as the explanation for the 
solar neutrino puzzle.

\subsection{Clues about the Higgs-Boson Mass}
\begin{figure}[tb] 
\centerline{\BoxedEPSF{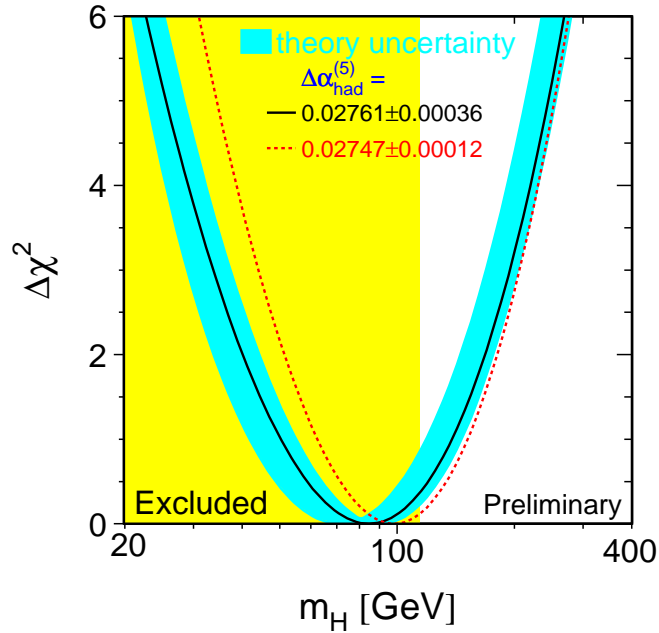 scaled 450}}
\vspace{-4pt}
\caption{$\Delta\chi^{2}=\chi^2-\chi^2_{\mathrm{min}}$ from a global fit to
	precision data {\it vs.} the Higgs-boson mass, $m_{H}$.  The solid
	line is the result of the fit; the band represents an estimate of the
	theoretical error due to missing higher order corrections.  The
	vertical band shows the 95\% CL exclusion limit on $m_{H}$ from the
	direct search at LEP. The dashed curve shows the sensitivity to a
	change in the evaluation of $\alpha(M_{Z}^{2})$.  (From the LEP
	Electroweak Working Group, Ref.\ {\protect\cite{ewwg}}.)}
\label{fig:blueband}
\end{figure}
Of particular interest in the realm of electroweak measurements is the
constraint on the mass of the Higgs boson. Figure~\ref{fig:blueband}
shows the $\Delta\chi^2$ curve derived from precision electroweak
measurements performed at LEP---and by SLD, CDF, D\O, NuTeV, and
others---as a function of the Higgs-boson mass, within the standard
model. The preferred value for the Higgs-boson mass, corresponding to
the minimum of the curve, is around $85\gevcc$, with an experimental
uncertainty of $+54$ and $-34\gevcc$ at 68\%\ confidence level. This
estimate is not materially affected by the NuTeV measurement discussed
above. 

Finding quantitative evidence that the virtual contributions of the
Higgs boson are required to describe the observed constellation of
electroweak data supports the theoretical contention that (something
like) the Higgs boson must exist.\footnote{Developed in \S5.1 of my
\textit{TASI~2000} lectures, Ref.~\cite{Quigg:2001td}.}
It also directs our attention for the search into a narrower range than
general theoretical arguments are capable of suggesting.
The precision electroweak measurements tell us that the mass of the
standard-model Higgs boson is lower than about $196\gevcc$ (one-sided 95\%
confidence level upper limit including both the experimental and the theoretical
uncertainty). 
Since the direct 
searches at LEP~\cite{unknown:2001xw} have concluded that $m_{H}> 114.1\gevcc$, excluding 
much of the favored region, either the Higgs boson is just around the 
corner, or the standard-model analysis is misleading.  One way or another,
things will soon be popping!

\section{Big Questions}
In organizing my thoughts about the future agenda of particle physics, I find
it useful to proceed from broad scientific themes to the specific questions---the
minute particulars---that will illuminate them,
and then to instruments and technology development. Here it will suffice to
cite the main themes and develop them briefly.

\vspace*{3pt}\noindent\textit{Elementarity.} Are the quarks and leptons structureless, or
will we find that they are composite particles with internal
structures that help us understand the properties of the individual
quarks and leptons?

\vspace*{3pt}\noindent\textit{Symmetry.} One of the most powerful lessons of the modern
synthesis of particle physics is that (local) symmetries prescribe
interactions.  Our investigation of symmetry must address the question
of which gauge symmetries exist (and, eventually, why).  We have
learned to seek symmetry in the laws of Nature, not necessarily in
the consequences of those laws.  Accordingly, we must understand how
the symmetries are hidden from us in the world we inhabit.  For the
moment, the most urgent problem in particle physics is to complete our
understanding of electroweak symmetry breaking by exploring the 1-TeV
scale.   With the passing of LEP2, this is the business of the experiments at the Tevatron
Collider and the Large Hadron Collider.

\vspace*{3pt}\noindent\textit{Unity.} In the sense of developing explanations that apply not
to one individual phenomenon in isolation, but to many phenomena in
common, unity is central to all of physics, and indeed to all of
science.  At this moment in particle physics, our
quest for unity takes several forms.

First, we have the fascinating possibility of gauge coupling
unification, the idea that all the interactions we encounter have a
common origin and thus a common strength at suitably high energy.

Second, there is the imperative of anomaly freedom in the electroweak
theory, which urges us to treat quarks and leptons together, not as
completely independent species.  Both of these ideas are embodied, of
course, in unified theories of the strong, weak, and electromagnetic
interactions, which imply the existence of still other forces---to
complete the grander gauge group of the unified theory---including
interactions that change quarks into leptons.

The third aspect of unity is the idea that the traditional distinction
between force particles and constituents might give way to a unified
understanding of all the particles.  The gluons of QCD carry color
charge, so we can imagine quarkless hadronic matter in the form of
glueballs.  Beyond that breaking down of the wall between messengers
and constituents, supersymmetry relates fermions and bosons.

Finally, we desire a reconciliation between the pervasive outsider,
gravity, and the forces that prevail in the quantum world of our
everyday laboratory experience.

\vspace*{3pt}\noindent\textit{Identity.} We do not understand the physics that sets quark
masses and mixings.  Although we are testing the idea that the phase
in the quark-mixing matrix lies behind the observed \textsf{CP}
violation, we do not know what determines that phase.  The
accumulating evidence for neutrino oscillations presents us with a new
embodiment of these puzzles in the lepton sector.  At bottom, the
question of identity is very simple to state: What makes an electron
and electron, and a top quark a top quark?

\vspace*{3pt}\noindent\textit{Topography.} ``What is the
dimensionality of spacetime?''  tests our preconceptions and unspoken
assumptions.  It is given immediacy by recent theoretical work.  For
its internal consistency, string theory requires an additional six or
seven space dimensions, beyond the $3+1$ dimensions of everyday
experience.  Until recently it has been presumed that the extra
dimensions must be compactified on the Planck scale, with a
stupendously small compactification radius $R \simeq
M_{\mathrm{Planck}}^{-1} = 1.6 \times 10^{-35}\m$.
Part of the vision of string theory is that what goes on in even such
tiny curled-up dimensions does affect the everyday world: excitations
of the Calabi--Yau manifolds determine the fermion
spectrum.

We have come to understand lately that Planck-length compactification is 
not---according to what we can establish---obligatory, and that current
experiment and observation admit the possibility of dimensions not 
navigated by the strong, weak, and electromagnetic interactions that 
are almost palpably large.  A whole range of new experiments will help 
us explore the fabric of space and time, in ways we didn't expect just 
a few years ago~\cite{smaria}.

\section{The Role of Gravity}
\subsection{The Vacuum Energy Problem}
I want to spend a moment to revisit a 
longstanding, but usually unspoken, challenge to the completeness of 
the electroweak theory as we have defined it: the vacuum energy 
problem \cite{Veltman:1975au}.
I do so not only for its intrinsic interest, but also to 
raise the question, ``Which problems of completeness and 
consistency do we worry about at a given moment?''  It is perfectly 
acceptable science---indeed, it is often essential---to put certain 
problems aside, in the expectation that we will return to them at the 
right moment.  What is important is never to forget that the problems 
are there, even if we do not allow them to paralyze us.  

For the usual Higgs potential, 
$V(\varphi^{\dagger}\varphi) = \mu^{2}(\varphi^{\dagger}\varphi) + 
\abs{\lambda}(\varphi^{\dagger}\varphi)^{2}$, the value of 
the potential at the minimum is
\begin{equation}
    V(\vev{\varphi^{\dagger}\varphi}) = \frac{\mu^{2}v^{2}}{4} = 
    - \frac{\abs{\lambda}v^{4}}{4} < 0.
    \label{minpot}
\end{equation}
Identifying $m_{H}^{2} = -2\mu^{2}$, we see that the Higgs potential 
contributes a field-independent constant term,
\begin{equation}
    \varrho_{H} \equiv \frac{m_{H}^{2}v^{2}}{8}.
    \label{eq:rhoH}
\end{equation}
I have chosen the notation $\varrho_{H}$ because the constant term in
the Lagrangian plays the role of a vacuum energy density.  When we
consider gravitation, adding a vacuum energy density
$\varrho_{\mathrm{vac}}$ is equivalent to adding a cosmological
constant term to Einstein's equation.  Although recent
observations\footnote{For a cogent summary of current knowledge of the
cosmological parameters, including evidence for a cosmological
constant, see Ref.\ \cite{cosconst}.} raise the intriguing possibility
that the cosmological constant may be different from zero, the
essential observational fact is that the vacuum energy density must be
very tiny indeed,\footnote{For a useful summary of gravitational
theory, see the essay by T.  d'Amour in \S14 of the 2000 \textit{Review
of Particle Physics,} Ref.~\cite{Groom:in}.}
\begin{equation}
    \varrho_{\mathrm{vac}} \ltap 10^{-46}\gev^{4}\; .
    \label{eq:rhovaclim}
\end{equation}
Therein lies the puzzle: if we take
$v = (G_F\sqrt{2})^{-\frac{1}{2}}  \approx 246\gev$  
and insert the current experimental lower bound 
\cite{unknown:2001xw}
$m_{H} \gtap 114\gevcc$ into \eqn{eq:rhoH}, we find that the 
contribution of the Higgs field to the vacuum energy density is
\begin{equation}
    \varrho_{H} \gtap  10^{8}\gev^{4},
    \label{eq:rhoHval}
\end{equation}
some 54 orders of magnitude larger than the upper bound inferred from 
the cosmological constant.

What are we to make of this mismatch?  The fact that $\varrho_{H} \gg 
\varrho_{\mathrm{vac}}$ means that the smallness of the cosmological 
constant needs to be explained.  In a unified theory of the strong, 
weak, and electromagnetic interactions, other (heavy!) Higgs fields 
have nonzero vacuum expectation values that may give rise to still 
greater mismatches.  At a fundamental level, we can therefore conclude 
that a spontaneously broken gauge theory of the strong, weak, and 
electromagnetic interactions---or merely of the electroweak 
interactions---cannot be complete.  Either we must find a separate 
principle to zero the vacuum energy density of the Higgs field, or 
we may suppose that a proper quantum theory of gravity, in combination 
with the other interactions, will resolve the puzzle of the 
cosmological constant.  The vacuum energy problem must be an important 
clue.  But to what?

\subsection{Neglecting Gravity}
It is entirely natural to neglect gravity in most particle-physics 
applications, because the coupling of a graviton $\mathcal{G}$ to a particle is 
tiny, generically of order $(E/M_{\mathrm{Planck}})$ where $E$ is a 
typical energy scale of the problem.  Thus, for example, we expect 
the branching fraction $B(K \rightarrow \pi\mathcal{G}) \sim 
(M_{K}/M_{\mathrm{Planck}})^{2} \sim 10^{-38}$.  And yet we cannot 
put gravity entirely out of our minds, even if we restrict our 
attention to standard-model interactions at attainable energies.

The great gap between the electroweak scale of about $10^{3}\gev$ and
the Planck scale of about $10^{19}\gev$ gives rise to the hierarchy
problem of the  electroweak theory \cite{hier}: how to protect the
Higgs-boson mass from quantum corrections that explore energies up to
$M_{\mathrm{Planck}}$. The conventional approach to the hierarchy
problem has been to ask why the electroweak scale (and the mass of the
Higgs boson) is so much smaller than the Planck scale.  Framing the
issue this way leads us to change the electroweak theory to include
supersymmetry~\cite{Allanach:2002nj}, or
technicolor~\cite{Hill:2002ap}, or some other
extension~\cite{Branson:2001pj,Carena:2002rm}. Supersymmetry posits one
or more new \textit{quantum dimensions,} and entails a fermion-boson
doubling of the particle spectrum. Superpartners on the 1-TeV scale
resolve the hierarchy problem by balancing the contributions of fermion
and boson loops to the Higgs-boson mass renormalization. Models
descended from technicolor ascribe the breaking of electroweak symmetry
to a new strong interaction that becomes strong on the weak scale,
making a composite Higgs boson. Many models developed over the past
decade accord a privileged role to the top quark and third generation.

Over the past few years, we have begun
instead to ask why gravity is so weak.  This question motivates us to
consider \textit{changing gravity} to understand why the Planck scale
is so large \cite{EDbiblio}.  Now,
elegant experiments that study details of Casimir and van der Waals
forces tell us that gravitation closely follows the Newtonian force
law down to distances on the order of $0.3\mm$~\cite{Hoyle:2000cv}, 
which corresponds to an energy scale of only about $10^{-12}\gev$! 
At shorter distances (higher energies), the constraints on deviations from Newton's
inverse-square force law deteriorate rapidly, so nothing prevents us
from considering changes to gravity even on a small but macroscopic
scale.

One way to change the force law is to imagine that gravity can
propagate into extra dimensions.  To respect the stronger constraints
on the behavior of the standard-model interactions, we suppose that
the $SU(3)_{c}\otimes SU(2)_{L}\otimes U(1)_{Y}$ gauge fields, plus
needed extensions, reside on $(3+1)$-dimensional branes, not in the
extra dimensions.

What difference do extra dimensions make?  The dimensional 
analysis (Gauss's law, if you like) that relates Newton's constant to 
the Planck scale changes.  If gravity propagates in $n$ extra 
dimensions with radius $R$, then
\begin{displaymath}
    G_{\mathrm{Newton}} \sim M_{\mathrm{Planck}}^{-2} \sim M^{\star\,-n-2}R^{-n}\; ,
    \label{eq:gauss}
\end{displaymath}
where $M^{\star}$ is gravity's true scale.  Notice that if we boldly 
take $M^{\star}$ to be as small as $1\tevcc$, then the radius of the extra 
dimensions is required to be smaller than about $1\mm$, for $n \ge 
2$.  If we use the four-dimensional force law to extrapolate the 
strength of gravity from low energies to high, we find that gravity 
becomes as strong as the other forces on the Planck scale.  If the force law 
changes at an energy $1/R$, as the large-extra-dimensions scenario 
suggests, then the forces are unified at lower energy $M^{\star}$.
What we know as the Planck scale is then a mirage that results 
from a false extrapolation: treating gravity as four-dimensional down 
to arbitrarily small distances, when in fact---or at least in this 
particular fiction---gravity propagates in $3+n$ spatial dimensions.  
The Planck mass is an artifact, given by $M_{\mathrm{Planck}} = 
M^{\star}(M^{\star}R)^{n/2}$.   If the true scale of gravity were close 
to $m_{H}$, the hierarchy problem would recede.

``Large'' extra dimensions present us with new ways to think about the
exponential variation of the Yukawa couplings that determine fermion masses.
If the standard-model brane
has a small thickness, the wave packets representing different fermion
species might have different locations within the extra
dimension~\cite{Arkani-Hamed:1999dc}.  On this
picture, the Yukawa couplings measure the overlap in the extra
dimensions of the left-handed and right-handed fermion wave packets and the
Higgs field, presumed pervasive.  Exponentially large differences
might then arise from small offsets in the new coordinate(s).  True or
not, it is a mind-expanding way to look at an important problem.

\section{Concluding Remarks}
In the midst of a revolution in our conception of Nature, we confront
many fundamental questions about our world of diversity and change. 
Are the quarks and leptons elementary or composite?  What are the
symmetries of Nature, and how are they hidden from us?  Will we find
new forms of matter, like the superpartners suggested by
supersymmetry?  Will we find additional fundamental forces?  What
makes an electron an electron and a top quark a top quark?  What is
the dimensionality of spacetime,  what is its shape?

These are themselves great questions and, in the usual way of science,
answering them can lead us toward the answers to yet broader and more
cosmic questions.  I believe that we are on the threshold of a
remarkable flowering of experimental particle physics, and of
theoretical physics that engages with experiment. Over the next decade
or two, we may hope to 
\begin{quote}
Understand electroweak symmetry breaking,
		Observe the Higgs boson,
		Measure neutrino masses and mixings,
		Establish Majorana neutrinos through the observation of neutrinoless double-beta decay,
		Thoroughly explore \textsf{CP} violation in $B$ decays,
		Exploit rare decays ($K$, $D$, \ldots),
		Observe the neutron's permanent electric dipole meoment, and pursue the electron's
		electric dipole moment,
		Use top as a tool,
		Observe new phases of matter,
		Understand hadron structure quantitatively,
		Uncover the full implications of QCD,
		Observe proton decay,
		Understand the baryon excess of the universe,
		Catalogue the matter and energy of the universe,
		Measure the equation of state of the dark energy,
		Search for new macroscopic forces,
		Determine the gauge symmetry that unifies the strong, weak, and electromagnetic interactions,
	Detect neutrinos from the universe,
	Learn how to quantize gravity,
		Learn why empty space is nearly weightless,
		Test the inflation hypothesis,
		Understand discrete symmetry violation,
		Resolve the hierarchy problem,
		Discover new gauge forces,
		Directly detect dark-matter particles,
		Explore extra spatial dimensions,
		Understand the origin of the large-scale structure of the universe,
		Observe gravitational radiation,
		Solve the strong \textsf{CP} problem,
		Learn whether supersymmetry operates on the TeV scale,
		Seek TeV-scale dynamical symmetry breaking,
		Search for new strong dynamics,
		Explain the highest-energy cosmic rays,
		Formulate the problem of identity, \ldots
\end{quote}
\noindent		
\ldots and learn the right questions to ask.\footnote{The vigorous discussion following my
talk raised a number of other fascinating issues that will surely be much on our minds---whether
we have the means to answer them soon, or not!}

As we contemplate the interplay between experiment and theory that will lead us to a new,
far-reaching understanding, it is inspiring to remember the words Michael
Faraday recorded in his
\textit{Research Notes} of 19th March 1849:
\begin{quote}
    ``Nothing is too wonderful to be true,\\ if it be consistent with the laws 
    of nature \ldots \\
    Experiment is the best test \ldots''
\end{quote}

Before closing, I'd like to step back from the specific aspirations of particle physics
to recall the larger significance---the significance for human society---of what we do.
In his millennial column in the New York \textit{Times,} Anthony Lewis writes~\cite{tonyl},
\begin{quote}
``[T]here has been one transforming change over
this thousand years. It is the adoption of the
scientiÞc method: the commitment to
experiment, to test every hypothesis. But it is
broader than science. It is the open mind, the
willingness in all aspects of life to consider
possibilities other than the received truth.
It is openness to reason.''
\end{quote}
Niels Bohr characterized the goal of science as the gradual reduction of prejudice.
When we are at our best---when we are truest to our ideals---this spirit is what we
offer to science, and to humanity.

\vspace*{6pt}
\centerline{$\star$\quad$\star$\quad$\star$\quad$\star$\quad$\star$}
\vspace*{6pt}

\noindent
What will we find at the LHC?
Here is Cecil Frank Powell, recalling the wonders revealed in the first emulsions 
exposed on the Pic du Midi:
\begin{quote}
``It was as if, suddenly, we had broken into a walled
orchard, where protected trees had flourished and all
kinds of exotic fruits had ripened in great profusion.''
\end{quote}
I expect wonders no less astonishing in the first collisions from the
LHC. I look forward, with you, to breaking into a new walled orchard
and feasting our eyes---and our minds---on the exotic fruits we will find
there.

\section*{Acknowledgment}
It is a great pleasure to thank our hosts and organizers for the stimulating program
and excellent welcome in Sardinia. I'd especially like to thank Gino Saitta and Carlo
Bosio for their many contributions to the success of the workshop. I'm also grateful
to Roger Cashmore for CERN's generous support of my participation.


\begin{thebibliography}{99}
\bibitem{dava} Dava Sobel, \textit{Galileo's daughter : a historical memoir of
science, faith, and love} (Walker \& Co., New York,  1999).

\bibitem{fnews} Judy Jackson, ``\textit{Interactions:} Communicating
particle physics in the twenty-first century,'' \textit{FermiNews} \textbf{25}
(March 29, 2002) p.~6, available online at
\url{http://www.fnal.gov/pub/ferminews/ferminews02-03-29/p2.html}.

\bibitem{Quigg:1999xg} C. Quigg, \app{30}{2145}{99}, \hepph{9905369}.


\bibitem{Quigg:2001td} Chris Quigg,  
``The Electroweak Theory,''  in \textit{Flavor Physics for the
Millennium,} Proceedings of \textit{TASI~2000,}    edited by Jonathan L. Rosner (World Scientific,
Singapore, 2001), pp.~3-67, FERMILAB-CONF-01-001-T,
\url{http://fnalpubs.fnal.gov/archive/2001/conf/Conf-01-001-T.html}.


\bibitem{Sirlin:1999zc}
A. Sirlin, ``Ten Years of Precision Electroweak Physics,'' in {\it
Proc.  19th Intl.  Symp.  on Photon and Lepton Interactions at
High Energy LP99, } ed.  J.~A. Jaros and M.~E. Peskin, \textit{Int.\ J.\ Mod.\
Phys.\ A} {\bf 15S1} (2000) 398 [eConf {\bf C990809}, 398 (2000)],
	\hepph{9912227}, \url{http://www.slac.stanford.edu/econf/C990809/docs/sirlin.pdf}.

\bibitem{Swartz:1999xv} M. Swartz, ``Precision Electroweak Physics at
the $Z$,'' in {\it Proc.  19th Intl.  Symp.  on Photon and
Lepton Interactions at High Energy LP99, } ed.  J.~A. Jaros and M.~E.
Peskin, \textit{Int.\ J.\ Mod.\ Phys.\ A} {\bf 15S1} (2000) 307 [eConf {\bf
C990809}, 307 (2000)], \hepex{9912026},
\url{http://www.slac.stanford.edu/econf/C990809/docs/swartz.pdf}.

\bibitem{Charlton:2001am}
D. Charlton, ``Experimental tests of the standard model,'' \textit{to
appear in the proceedings of International Europhysics Conference on
High-Energy Physics (HEP 2001), Budapest, Hungary, 12-18 Jul 2001,}
\hepex{0110086}.


\bibitem{unknown:2001xw}
ALEPH Collaboration, DELPHI Collaboration, L3 Collaboration, OPAL
Collaboration, the LEP Higgs Working Group, ``Search for the Standard
Model Higgs Boson at LEP,'' \hepex{0107029}.


\bibitem{Quigg:1999di} C. Quigg, ``The State of the Standard Model,''
 in \textit{Physics Potential and Development of Muon
Colliders and Neutrino Factories,} edited by David B. Cline, AIP
Conference Proceedings 542 (American Institute of Physics, Melville,
NY, 2000), pp.~3--28, \hepph{0001145}.

\bibitem{Wilczek:1999id}
F. Wilczek,  \npM{A663}{3}{2000}.

\bibitem{sbt} S.~B. Treiman, {\em The Odd Quantum}
(Princeton University Press, Princeton, 1999).

\bibitem{ewwg} LEP Electroweak Working Group,
\url{http://www.cern.ch/LEPEWWG/.}

\bibitem{Zeller:2001hh}
G.~P. Zeller, \etal\  [NuTeV Collaboration],
\prll{88}{091802}{2002}.

\bibitem{Chanowitz:2001bv}
M.~S. Chanowitz,  \prll{87}{231802}{2001}, argues that,  combining
precision measurements and the Higgs boson search limit, the
electroweak data favor new physics whether the $A_{\mathrm{fb}}^{0,b}$
anomaly is genuine or not.


\bibitem{Wilczek:be}
F. Wilczek, \phystoday{52}{11}{11}{99}.


\bibitem{Aoki:1999yr} S. Aoki, \etal\ (CP--PACS Collaboration), 
\prll{84}{238}{2000}.

\bibitem{Buras:1977yy}
A.~J. Buras, J.~R. Ellis, M.~K. Gaillard, and D.~V. Nanopoulos,
\np{B135}{66}{78}. 

\bibitem{Toshito:2001dk}
T. Toshito  [Super-Kamiokande Collaboration],
``Super-Kamiokande atmospheric $\nu$ results,''
\hepex{0105023}.

\bibitem{Ahmad:2001an}
Q.~R. Ahmad, \etal\  [SNO Collaboration],
\prll{87}{071301}{2001}.


\bibitem{Fukuda:2001nj}
S. Fukuda, \etal\  [Super-Kamiokande Collaboration],
\prll{86}{5651}{2001}.

\bibitem{smaria} J.~Hewett and M.~Spiropulu, ``Particle-Physics Probes
of Extra Spacetime Dimensions,'' to appear in \textit{Annu. Rev. Nucl.
Part. Sci.} \textbf{52} (2002).

\bibitem{Veltman:1975au}
M. Veltman, 
\prl{34}{777}{75}; 
A.~D. Linde,
\textit{JETP Lett.} {\bf 19} (1974) 183; 
J. Dreitlein, 
\prl{33}{1243}{77}; 
S. Weinberg,
\rmp{61}{1}{89}.

\bibitem{cosconst} M.~S. Turner, ``Cosmological Parameters,'' in
\textit{COSMO--98: Second International Workshop on Particle Physics and 
the Early Universe,} edited by David O. Caldwell, AIP
Conference Proceedings 478 (American Institute of Physics, Woodbury,
NY, 1999) Woodbury, N.Y., Amer. Inst. Phys., 1999), p.~113,
\astro{9904051}.  

\bibitem{Groom:in}
D.~E. Groom, \etal\ (Particle Data Group),
\textit{Euro. Phys. J. C} \textbf{15} (2000) 1 and 2001
 off-year partial update for the 2002 edition at \url{http://pdg.lbl.gov/}
 
\bibitem{hier} E. Gildener, \pr{14}{1667}{76}; S. Weinberg, 
\pl{82B}{387}{79}.

\bibitem{Allanach:2002nj}
B.~C.~Allanach, \etal, ``The Snowmass points and slopes: Benchmarks for
SUSY searches,'' \hepph{0202233}.

\bibitem{Hill:2002ap}
C.~T. Hill and E.~H. Simmons, 
``Strong dynamics and electroweak symmetry breaking,''
\hepph{0203079}.

\bibitem{Branson:2001pj}
J.~G.~Branson, D.~Denegri, I.~Hinchliffe, F.~Gianotti, F.~E.~Paige and
P.~Sphicas [ATLAS Collaboration and CMS Collaboration], ``High
transverse momentum physics at the Large Hadron Collider,''
\hepph{0110021}.

\bibitem{Carena:2002rm} M.~Carena, D.~W.~Gerdes, H.~E.~Haber,
A.~S.~Turcot and P.~M.~Zerwas, ``Executive summary of the Snowmass 2001
working group (P1) \textit{electroweak  symmetry breaking,}''
\hepph{0203229}.




\bibitem{EDbiblio} Among the seminal papers, see I. Antoniadis, 
\pl{B246}{377}{90};
J.~D. Lykken,  \pr{54}{3693}{96};
N. Arkani-Hamed, S. Dimopoulos,  and G. Dvali, \pl{B429}{263}{98}.


\bibitem{Hoyle:2000cv} C.~D. Hoyle, \etal, \prll{86}{1418}{2001}.

\bibitem{Arkani-Hamed:1999dc}
N. Arkani-Hamed,  and M. Schmaltz, \prM{61}{033005}{2000};
E.~A. Mirabelli and M. Schmaltz, \prM{61}{113011}{2000}.

\bibitem{tonyl} Anthony Lewis, ``\textit{Abroad at Home:} The Fault, Dear Brutus,''
New York \textit{Times,} 31 December 1999.

\end{thebibliography}
\end{document}